\documentclass[11pt]{article}
\usepackage{graphics,wrapfig,graphicx} 
 
\begin{document} 
\title{Longitudinal Single-Spin Asymmetries in Proton-Proton Scattering 
with a Hadronic Final State}
 
\author{S.~Arnold$^{(1)}$, A.~Metz$^{(2)}$, 
 W.~Vogelsang$^{(3)}$
\\[.5cm]
(1) \it Institut f\"{u}r Theoretische Physik II, Ruhr-Universit\"{a}t, 
44780 Bochum, Germany 
\\\\ 
(2) \it Department of Physics, Temple University, Philadelphia, 
PA 19122, USA 
\\\\ 
(3) \it Physics Department, Brookhaven National Laboratory, Upton, NY
11973, USA} 
 
\date{\today} 
\maketitle 
 
\vspace*{-10cm}
\begin{flushright}
BNL-NT-08/20 \\
\end{flushright}
\vspace*{8cm}
\begin{abstract} 
We consider longitudinal, parity-violating single-spin asymmetries 
in proton-proton collisions at RHIC. The focus of this study is on the 
production of single-inclusive jets, as well as on jets that contain a 
charm quark. While the asymmetry for inclusive jets turns out to be 
small, we find considerably larger effects for the case of charm 
production. We also investigate the role of leading
threshold logarithms and find that they increase the polarized and
unpolarized cross sections and reduce the spin asymmetry. 
\end{abstract} 
 
\section{Introduction} 
Over the past decades already much information on the  
nucleon's quark and anti-quark helicity distributions, $\Delta q$ and 
$\Delta \bar{q}$, has been collected. 
The key processes exploited so far are inclusive and 
semi-inclusive deep-inelastic lepton nucleon scattering (DIS) 
(see Refs.~\cite{Airapetian:2007mh,Alekseev:2007vi} for recent experimental 
work). Inclusive DIS only measures the combinations $\Delta q+
\Delta \bar{q}$, while semi-inclusive DIS allows to separate the quark
and anti-quark distributions, albeit with relatively large uncertainties
associated with fragmentation. Entirely independent and complementary 
information on the $\Delta q$ and $\Delta \bar{q}$ will be obtained 
from the study of parity-violating single-spin asymmetries (SSAs)
in proton-proton scattering 
at RHIC~\cite{BouSof,Bunce:2000uv,Masterspin:2005}.

The main focus of the measurements of parity-violating spin asymmetries
planned at RHIC is on the production of $W$-bosons and their subsequent 
decay into a charged lepton and the corresponding unobserved (anti-)neutrino. 
This process can lead to large longitudinal SSAs, of the order of $50\,\%$.
Moreover, the process is ``clean'' in the sense that only the single 
generic subprocess $q\bar{q}'\to W\to l\nu$ contributes to lowest
perturbative order (LO), which in principle allows a direct study of a 
particular quark or anti-quark helicity 
distribution~\cite{BouSof,Bunce:2000uv,Masterspin:2005}. Also, as the 
final state is generated by the electroweak interaction,
QCD radiative corrections to the process are relatively simple to 
calculate and well understood~\cite{Nado}.
On the other hand, the counting rates for lepton final states are low.

In the present note, we discuss parity-violating SSAs with a hadronic  
final state, such as a jet at large transverse momentum $p_T$. This
potentially provides an interesting alternative to the leptonic signal,
foremost because jets are produced much more copiously than leptons.
The idea to use parity-violating SSAs to see hadronic $W$ 
decays~\cite{PTT} actually predates the discovery of the $W$s. The
specific application to RHIC with the aim of obtaining information 
on the spin-dependent quark and anti-quark distributions was
discussed in~\cite{Bourrely:1990pz}, where also the results for
all the associated LO subprocess cross sections were presented. 
First-order QCD corrections were computed in Ref.~\cite{Moretti:2005aa},
and turn out to be quite significant. Apart from updating the 
expectations for the LO parity-violating SSA by using a more recent
set of polarized parton distributions, we extend the theoretical 
status of higher-order QCD corrections by taking into account the 
resummation of large leading threshold logarithms to the cross
sections. We also discuss the special case of jets that contain 
a charm quark. We find that in this case a much more favorable SSA is 
obtained than for the single-inclusive jet case.

\section{Parity-violating longitudinal SSAs for single jet
production} 

A non-vanishing longitudinal
SSA in the process $\vec{p} + p \to \textrm{jet} + X$ is
parity-violating and can only arise through participation of 
the weak interactions. Here we consider this reaction at the lowest order of 
perturbation theory. All the involved partonic cross sections for the process 
have been known for a long time~\cite{Bourrely:1990pz}. We have confirmed the 
corresponding results. There are five different helicity structures 
for these cross sections, 
\vspace{0.2cm} \\ 
\begin{tabular}{lll} 
  (1) $\left(1-\lambda_1\lambda_2 \right)$ & (2) 
$\left(1+\lambda_1\lambda_2 \right)$&\\ 
  (3) $\left(1-\lambda_1 \right) \left(1-\lambda_2 \right)$ & 
  (4) $\left(1+\lambda_1 \right) \left(1+\lambda_2 \right)$ & 
  (5) $\left(1-\lambda_1 \right) \left(1+\lambda_2 \right)$ , \\ 
\end{tabular} 
\vspace{0.2cm} 
\\ 
where $\lambda_1$ and $\lambda_2$ denote the helicities of the 
partons in the initial state. All these structures contribute 
to the spin-averaged $pp$ cross section,
$\sigma_{{\mathrm{unp}}} = \left( \sigma_{++} + \sigma_{+-} 
 + \sigma_{-+} + \sigma_{--} \right)/4$,
while the parity-conserving structures (1) and (2) drop out in the 
polarized cross section, which we define according to
$\sigma_{{\mathrm{pol}}} = \left( \sigma_{++} + \sigma_{+-} 
 - \sigma_{-+} - \sigma_{--} \right)/4$,
the subscripts denoting the proton helicities. Note that this definition 
implies that the helicities of the ``second'' proton are summed over,
so that we are considering a $\vec{p}p$ collision, and we can also write 
$\sigma_{{\mathrm{pol}}} = \left( \sigma_{+} 
 - \sigma_{-}\right)/2$, where the helicities refer to the polarized
proton. Also note that in the following the rapidity of a produced 
final-state particle will be counted positive in the forward direction
of the ``first'' (the polarized) proton.  

In terms of coupling constants, three different types of processes are 
taken into account. First, the ${\cal O} (\alpha_S^2)$ pure QCD 
$\textrm{(2-parton)} \to \textrm{(2-parton)}$ processes. Because these 
do not violate parity, they only contribute to the unpolarized cross section.
Second, partonic processes of ${\cal O} (\alpha_S \alpha_W)$, i.e., 
interference terms between strong and electroweak amplitudes. These
generate a non-vanishing SSA~\cite{Bourrely:1990pz}.
The interference with the strong interaction generally leads to larger 
counting rates for the polarized cross section, compared to a reaction 
with a leptonic final state. Finally, there are purely electroweak 
partonic processes of ${\cal O} (\alpha_W^2)$. 
While these are of lesser importance for the asymmetry for jet production,
they dominate in the case of charm production, as we will discuss below.
All in all, the asymmetry is schematically of the form 
$\left[{\cal O} (\alpha_S \alpha_W) + {\cal O} (\alpha_W^2)\right]/
\left[{\cal O}(\alpha_S^2) + {\cal O}(\alpha_S \alpha_W) + 
{\cal O}(\alpha_W^2)\right]$. Regardless of the fact that the various
terms have mixed perturbative orders in $\alpha_S$ and $\alpha_W$,
we collectively refer to them as ``leading order'', since they all contribute
at $2\to 2$ tree level.

The spin-averaged hadronic jet production cross section is given by 
\begin{equation}\label{e:sigma}
\frac{d\sigma_{{\mathrm{unp}}}}{d p_T d\eta} =  \sum_{a,b} 
\frac{1}{1+\delta_{ab}}\int_{x_{{\mathrm{min}}}}^1 
d x_1 \frac{2 p_T}{x_1 - \frac{p_T}{\sqrt{s}}e^{\eta}} \, 
\Big[ x_1 f^{a/p}(x_1,\mu^2) \, x_2 f^{b/p}(x_2,\mu^2) \, 
\frac{d\hat{\sigma}_{ab,{\mathrm{unp}}}}{d\hat{t}} + 
(a \leftrightarrow b) \Big] \; ,
\end{equation}
where $p_T$ and $\eta$ are the jet's transverse momentum and pseudo-rapidity,
respectively, and where $x_{{\mathrm{min}}} = x_T {\mathrm{e}}^{\eta}/
(2 - x_T {\mathrm{e}}^{-\eta})$, $x_2= x_1 x_T {\mathrm{e}}^{-\eta}/
(2 x_1 - x_T {\mathrm{e}}^{\eta})$ with $x_T=2 p_T/\sqrt{s}$. 
The $d\hat{\sigma}_{ab,{\mathrm{unp}}}/
d\hat{t}$ with $\hat{t}=(p_a-p_{\mathrm{jet}})^2$ are  the spin-averaged 
$2\to 2$ partonic cross sections, and the $f^{a,b/p}$ the distributions 
for partons $a,b$ in the proton.
To compute $\sigma_{{\mathrm{pol}}}$ one just has to use the appropriate
polarized partonic cross sections in Eq.~(\ref{e:sigma}), and to replace the 
distribution $f^{a/p}$ by the corresponding helicity distribution 
$\Delta f^{a/p}$. In our numerical calculations we use the
parton distribution functions of~\cite{Gluck:1998xa,Gluck:2000dy}. We
choose $\mu = p_T$ for the scale of the parton densities and of the 
strong coupling constant. 
\begin{figure}[t] 
\begin{center} 
\hspace*{-4mm}
\begin{tabular}{cc} 
\begin{minipage}{0.5\columnwidth} 
\resizebox{0.99\columnwidth}{!}{\includegraphics[50,50][410,302]{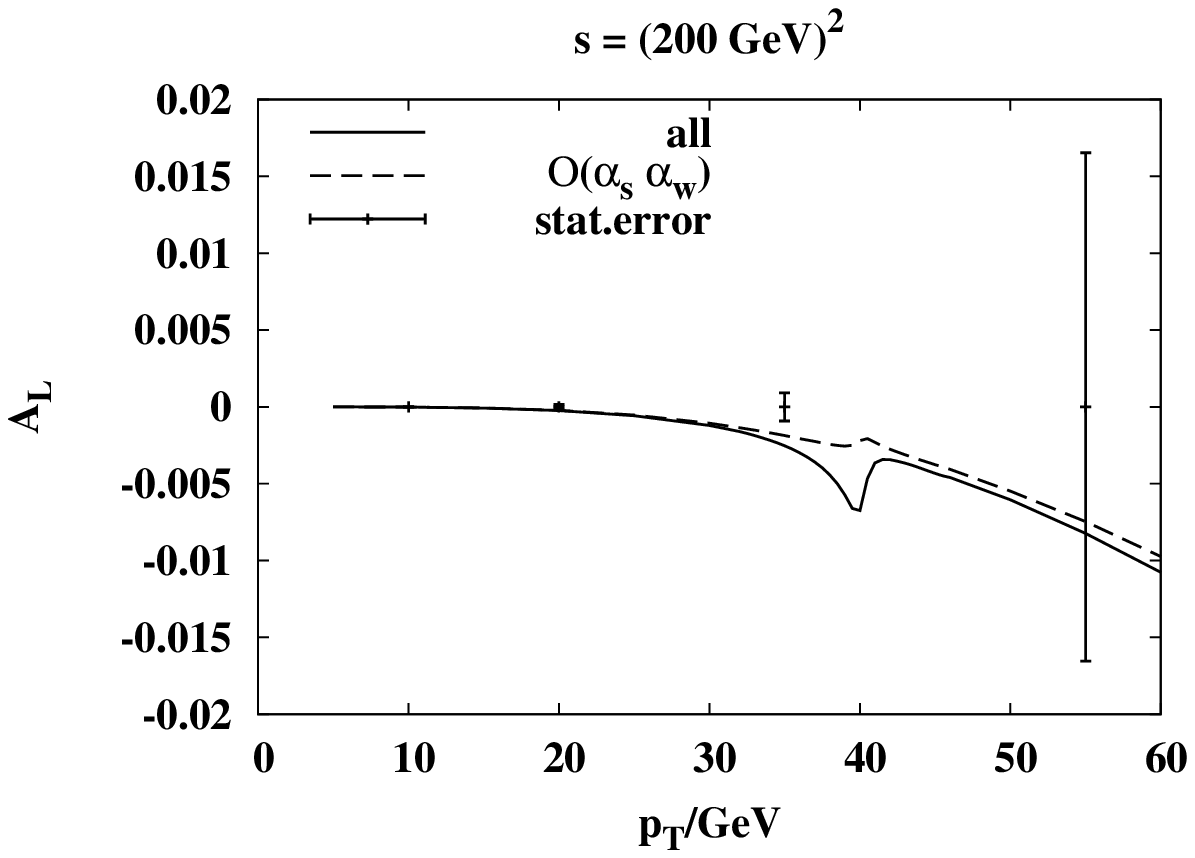}} 
\centerline{(a)} 
\end{minipage} & 
        
\begin{minipage}{0.5\columnwidth} 
\resizebox{0.99\columnwidth}{!}{\includegraphics[50,50][410,302]{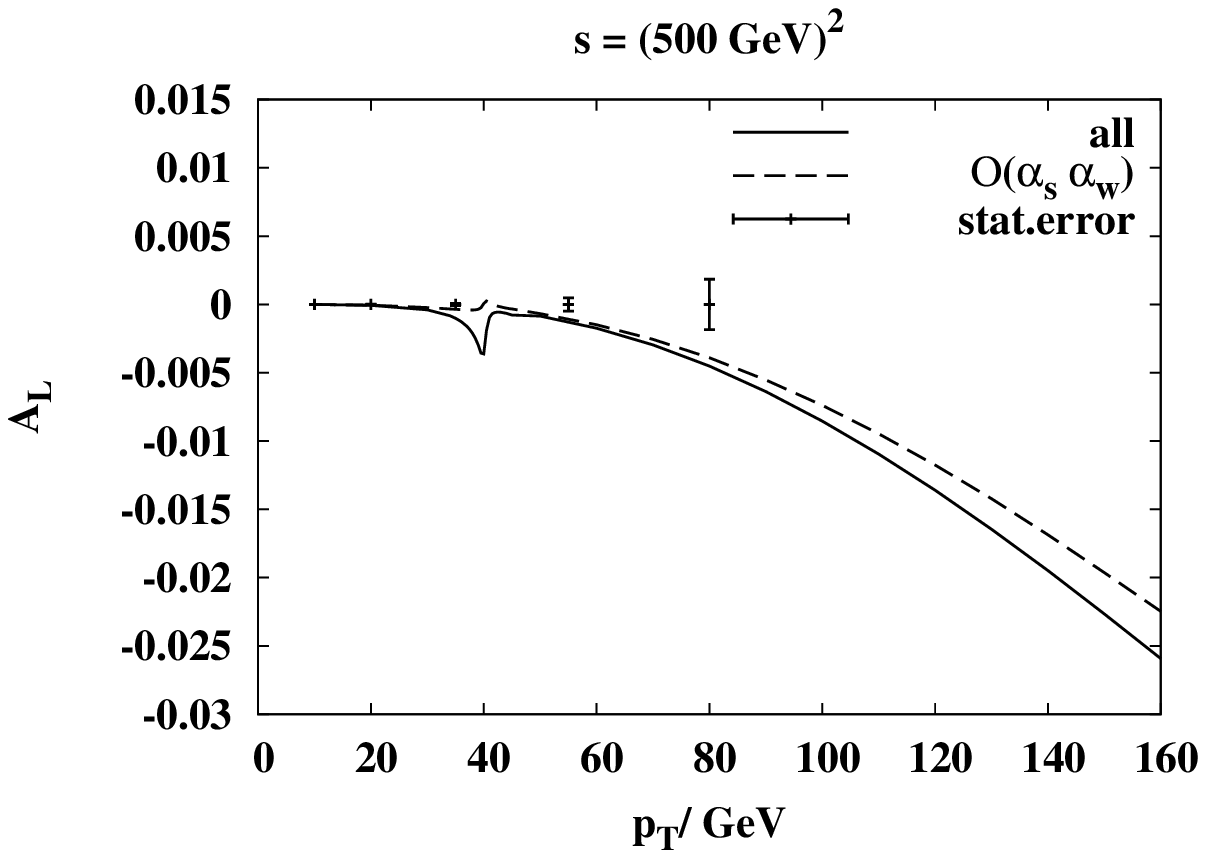}}
\centerline{(b)} 
\end{minipage} \\ 
\end{tabular} 
\caption{The single-spin asymmetry $A_L$ for single-inclusive jet production,
for the two RHIC energies. Dashed lines: contributions of 
${\cal O}(\alpha_S^2)+ {\cal O}(\alpha_S\alpha_W)$; solid lines: 
including also the contributions of ${\cal O}(\alpha_W^2)$. 
The indicated statistical error projections are estimated assuming an integrated 
luminosity of $100 \, \textrm{pb}^{-1}$ for $\sqrt{s} = 200 \, \textrm{GeV}$
and of $500 \, \textrm{pb}^{-1}$ for $\sqrt{s} = 500 \, \textrm{GeV}$, as well
as a beam polarization of $70\%$.} 
\label{f:jet} 
\end{center} 
\end{figure} 

\begin{wrapfigure}{l}[0cm]{0.5\textwidth}
\includegraphics[width=0.48\textwidth]{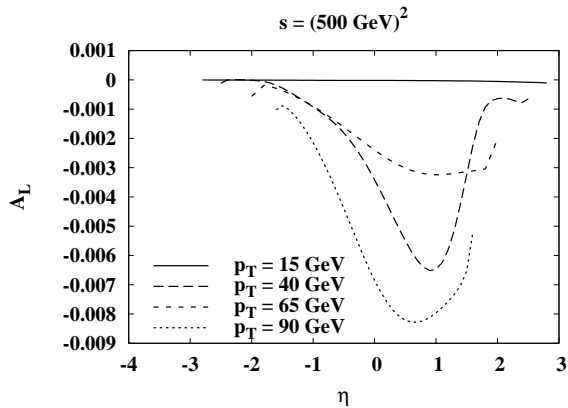}
\caption{$A_L$ for jet production at $\sqrt{s} = 500 \, \textrm{GeV}$ as a 
function of $\eta$ for different values of $p_T$.}
\label{f:jet_eta}
\end{wrapfigure} 

Figure~\ref{f:jet} shows the spin asymmetry $A_L = 
\sigma_{{\mathrm{pol}}}/\sigma_{{\mathrm{unp}}}$ for two 
different center-of-mass (c.m.) energies, 
$\sqrt{s} = 200 \, \textrm{GeV}$, the present RHIC energy, and 
$\sqrt{s} = 500 \, \textrm{GeV}$, the energy that will be
used for the $W$ physics program. We present $A_L$ as a function of 
$p_T$, integrated over $-1<\eta<1$. For the solid lines in Fig.~\ref{f:jet} 
all partonic contributions have been included, while for the dashed lines
the purely electroweak contributions of ${\cal O}(\alpha_W^2)$ have been
excluded. $A_L$ increases with $p_T$, but it remains overall rather small. 
This is due to the large QCD contribution in the denominator of the asymmetry,
which has gluon-induced contributions and also the larger 
(and positive-definite) partonic cross sections. 
Purely electroweak contributions are only important starting from 
$p_T \simeq 30-40 \,\textrm{GeV}$, where the (resonant) $s$-channel production 
of weak gauge bosons becomes large.
In Fig.~\ref{f:jet} we have also displayed the statistical errors to be 
expected for integrated luminosity ${\cal L} = 100 \, \textrm{pb}^{-1}$ for
$\sqrt{s} = 200 \, \textrm{GeV}$ and ${\cal L} = 500 \, \textrm{pb}^{-1}$ for 
$\sqrt{s} = 500 \, \textrm{GeV}$, as well as a beam polarization $P=70\,\%$. 
These have been estimated as
$\delta A_L =1/(P\sqrt{{\cal L} \cdot \sigma_{{\mathrm{unp}}}})$,
where $\sigma_{{\mathrm{unp}}}$ has been integrated over the 
following $p_T$-bins: (5-15), (15-25), (25-45), (45-65), (65-95)$\,$GeV. Even though the asymmetry is very small, it may 
nevertheless be measurable at $\sqrt{s}=500$~GeV, at high $p_T$.

Figure~\ref{f:jet_eta} shows $A_L$ for several fixed $p_T$ values 
as a function of $\eta$. $|A_L|$ becomes maximal 
at positive 
$\eta$ when a large-$x$ quark from the polarized proton participates
in the scattering. The plot shows that by integrating over a 
suitable $\eta$-range one may optimize between magnitude of the 
asymmetry on the one hand and the size of the statistical error bars on 
the other. Measurements at large positive $\eta$ might
give valuable information on the helicity distributions at 
large $x$, complementary to that obtained in lepton nucleon
scattering.

\section{Resummation of leading threshold logarithms} 
Near the partonic threshold, $\hat{x}_T = 2p_T/\sqrt{\hat{s}}\sim 1$,
when the initial partons have just enough energy to produce the 
high-$p_T$ jet and an unobserved recoiling partonic
final state, large logarithmic corrections of the form  
$\alpha_S^k \ln^{m}\left(1 - \hat{x}_T^2 \right)$ with 
$m\leq 2k$ arise at the $k$th order of perturbation theory. 
These result from the emission of soft and collinear gluons.
If the threshold region plays a significant role for the hadronic
cross section, which is the case for RHIC when the jet transverse momentum 
becomes large, the logarithmic corrections need to be resummed to 
all orders in $\alpha_S$, at least for the leading towers of logarithms, 
in order to maintain a useful perturbative expansion.
The resummation is usually performed in Mellin-$N$ moment space, 
with moments taken in $x_T^2$. To leading double logarithmic (LL) accuracy 
($m= 2k$), which we will focus on in the present study, the resummed 
cross section is for each partonic subprocess given 
by~\cite{deFlorian:2007fv} 
\begin{equation} \label{e:resum}
\hat{\sigma}_{ab}^{{\mathrm{(res)}}}(N) = \sum_{c,d} \, \Delta_N^a \, 
\Delta_N^b \, J_N^d \,
\hat{\sigma}^{{\mathrm{(Born)}}}_{ab\rightarrow cd}(N) \, ,
\end{equation}  
where $\hat{\sigma}^{{\mathrm{(Born)}}}_{ab\rightarrow cd}(N)$
are the moments of the Born cross sections. Here, parton $c$ is 
the ``observed'' final-state parton that produces the jet. 
$\Delta_N^{a,b}$ and $J_N^d$ are ``radiative factors'' associated
with gluon emission from the external legs $a,b,d$ of the Born process. 
They contain all leading logarithms. As was discussed 
in~\cite{deFlorian:2007fv}, emission off parton $c$ does not produce 
double logarithms. For details and the precise definition of the 
various factors in Eq.~(\ref{e:resum}), see~\cite{deFlorian:2007fv} 
and the references therein.

Eq.~(\ref{e:resum}) as written applies to the cross section integrated
over all $\eta$, which takes the form of a genuine mathematical 
convolution of parton distributions and partonic cross sections. 
The Mellin moments of the hadronic cross section in $x_T^2$ can 
be computed from this as $\sigma(N) = \sum_{a,b}  
\hat{\sigma}_{ab}^{\mathrm{(res)}} (N)f^{a/p}(N)f^{b/p}(N)$, where 
$f^{a/p}(N),f^{b/p}(N)$ are the Mellin moments of the parton densities. 
The hadronic cross section as a function of $p_T$
would then be obtained by applying an inverse Mellin transformation.
However, in the LL approximation, one can follow a somewhat simpler 
procedure. We can absorb the resummation effects into the parton 
distribution functions in moment space, multiplying for example 
$f^{a/p}(N)$ by the associated radiative factor $\Delta_N^a$, 
and likewise for parton $b$. One next also includes the factor $J_N^d$
appropriately. To give a specific example, let us consider the 
subprocess $qq\to qq$. We write the resummed cross section as
\begin{equation} \label{e:absorb}
\hat{\sigma}_{qq}^{{\mathrm{(res)}}}(N)=
\underbrace{[ f^{q/p}(N) \, \Delta_N^q ]}_{\to \, \tilde{f}^{q/p} (x_1)} \; 
\underbrace{[ f^{q/p}(N) \, \Delta_N^q \, J_N^q ]}_{\to \, \hat{f}^{q/p} 
(x_2)} \;  
\hat{\sigma}^{{\mathrm{(Born)}}}_{qq\to qq} \,.
\end{equation} 
We have introduced in this equation ``pseudo''-parton distributions 
$\tilde{f}^{q/p}$ and $\hat{f}^{q/p}$ that incorporate the resummation 
effects. Use of these in the ordinary momentum-space LO cross sections 
of~\cite{Bourrely:1990pz} will evidently give a result equivalent to the
full LL resummation in Mellin space, but with the advantage that one 
does not need to compute the Mellin moments of the Born cross sections.
Of course, for a general partonic 
process there is an ambiguity regarding whether one absorbs the
factor $J_N^{q,g}$ into $f^{a/p}$ or $f^{b/p}$. To compensate for this, 
we simply symmetrize afterwards by using the combination  
$(\tilde{f}^{a/p} (x_1) \, \hat{f}^{b/p} (x_2)  + 
\hat{f}^{a/p} (x_1) \, \tilde{f}^{b/p} (x_2))/2$. This plays no 
role if the cross section is integrated over all $\eta$. If the integration
over $\eta$ is performed only over a limited (but large and symmetric) range, 
the above procedure is no longer exact, but should still give a good 
approximation to the expected LL resummation effects. 

\begin{figure}[t] 
\begin{center} 
\hspace*{-5mm}
\begin{tabular}{cc} 
\begin{minipage}{0.5\columnwidth} 
\resizebox{0.99\columnwidth}{!}
{\includegraphics[50,50][410,302]{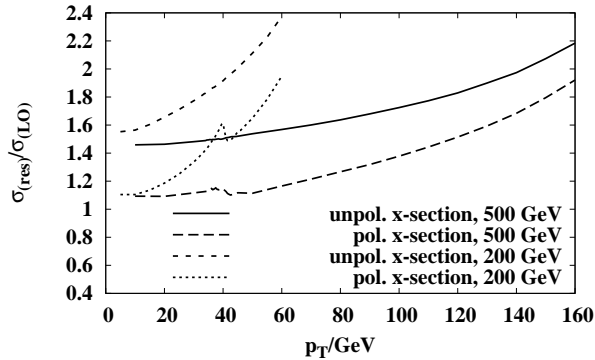}} \end{minipage} & 
\end{tabular} 
\caption{LL threshold resummation effects for single-inclusive
jet production. The plots show the 
ratios $\sigma_{{\mathrm{unp}}}^{{\mathrm{(res)}}}/
\sigma_{{\mathrm{unp}}}^{{\mathrm{(LO)}}}$ and
$\sigma_{{\mathrm{pol}}}^{{\mathrm{(res)}}}
/\sigma_{{\mathrm{pol}}}^{{\mathrm{(LO)}}}$ for two different 
c.m.-energies.} 
\label{f:resum} 
\end{center} 
\end{figure}

Figure~\ref{f:resum} shows our results for the LL threshold resummation 
for the unpolarized and polarized cross sections. One can see that 
in general resummation increases the jet cross sections at RHIC, as 
expected~\cite{deFlorian:2007fv}. It turns out that the unpolarized 
cross section is more affected by resummation than the polarized one.
This is also expected since processes with gluon initial states, 
which typically receive the largest resummation effects due to the
larger color charge of gluons, are absent in the polarized case at LO.
As a result, resummation tends to further decrease the parity-violating 
asymmetry $A_L$, by roughly $30\,\%$. This finding is qualitatively 
consistent with the results of the full NLO calculation 
of~\cite{Moretti:2005aa}, even though more detailed studies are
needed here.  

\section{Charm production}

\begin{wrapfigure}{l}[0cm]{0.5\textwidth}
\includegraphics[width=0.48\textwidth]{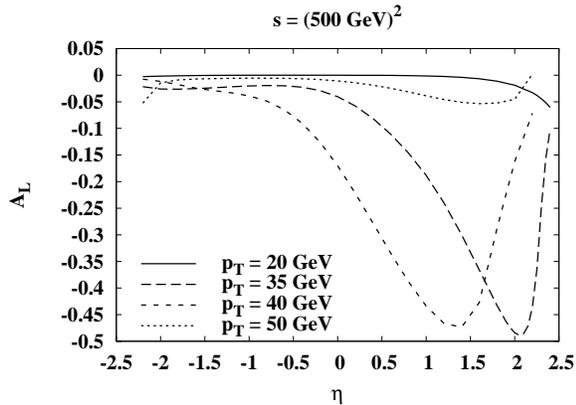}
\caption{$A_L$ for charm production at $\sqrt{s} = 500 \, \textrm{GeV}$  
as a function of $\eta$, for different values of $p_T$.}
\label{f:charm_eta}
\vspace*{-6mm}
\end{wrapfigure}

One can strongly reduce the large gluon-induced QCD background that
suppresses $A_L$ if one considers an SSA for the production of a 
charm quark, i.e., the process $\vec{p} + p \to c + X$. Since we take
the charm quark to be at high transverse momentum, we can neglect its mass.  
In order to have a realistic calculation, one would need to include a
fragmentation function for the charm quark turning into a charmed hadron.
For our first exploratory study here, however, we neglect the fragmentation
process.
\begin{figure}[t] 
\begin{center} 
\hspace*{-4mm}
\begin{tabular}{cc} 
\begin{minipage}{0.5\columnwidth} 
\resizebox{0.99\columnwidth}{!}
{\includegraphics[50,50][410,302]{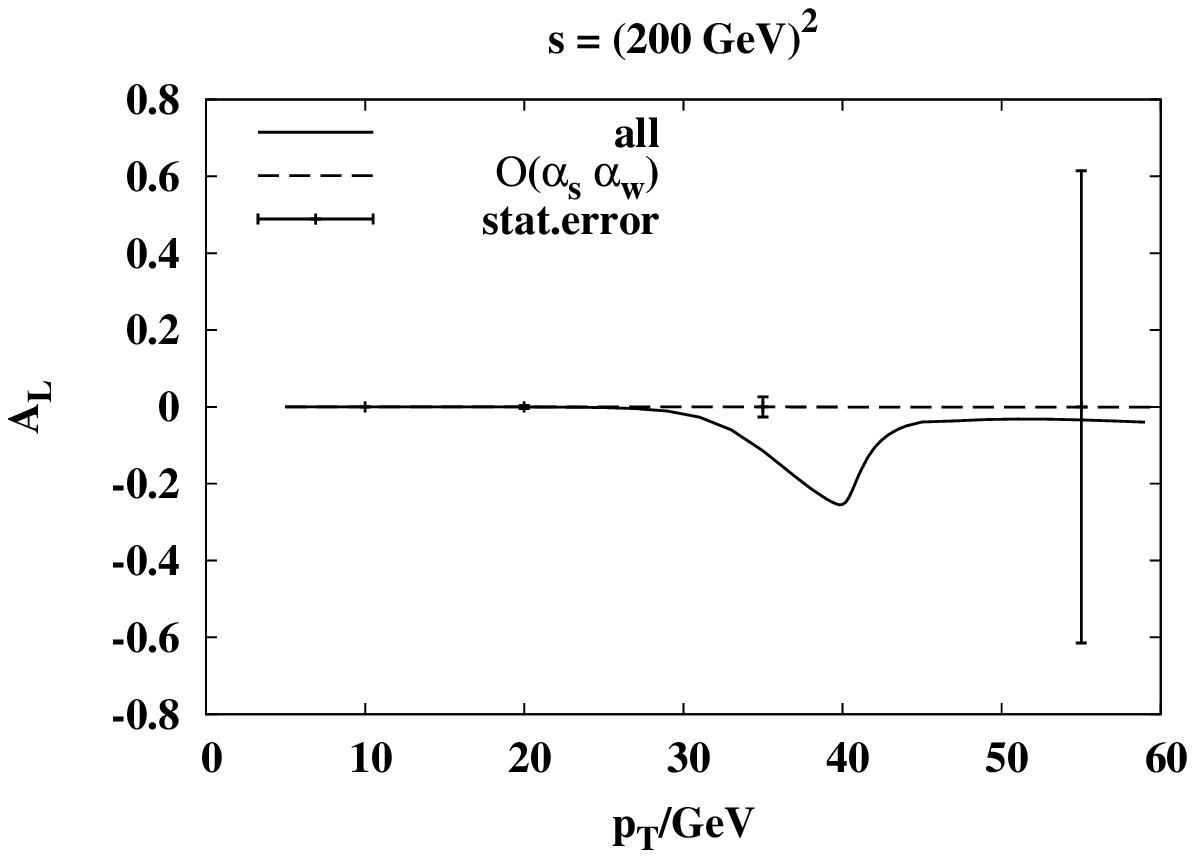}}
\centerline{(a)} 
\end{minipage} & 
\begin{minipage}{0.5\columnwidth} 
\resizebox{0.99\columnwidth}{!}
{\includegraphics[50,50][410,302]{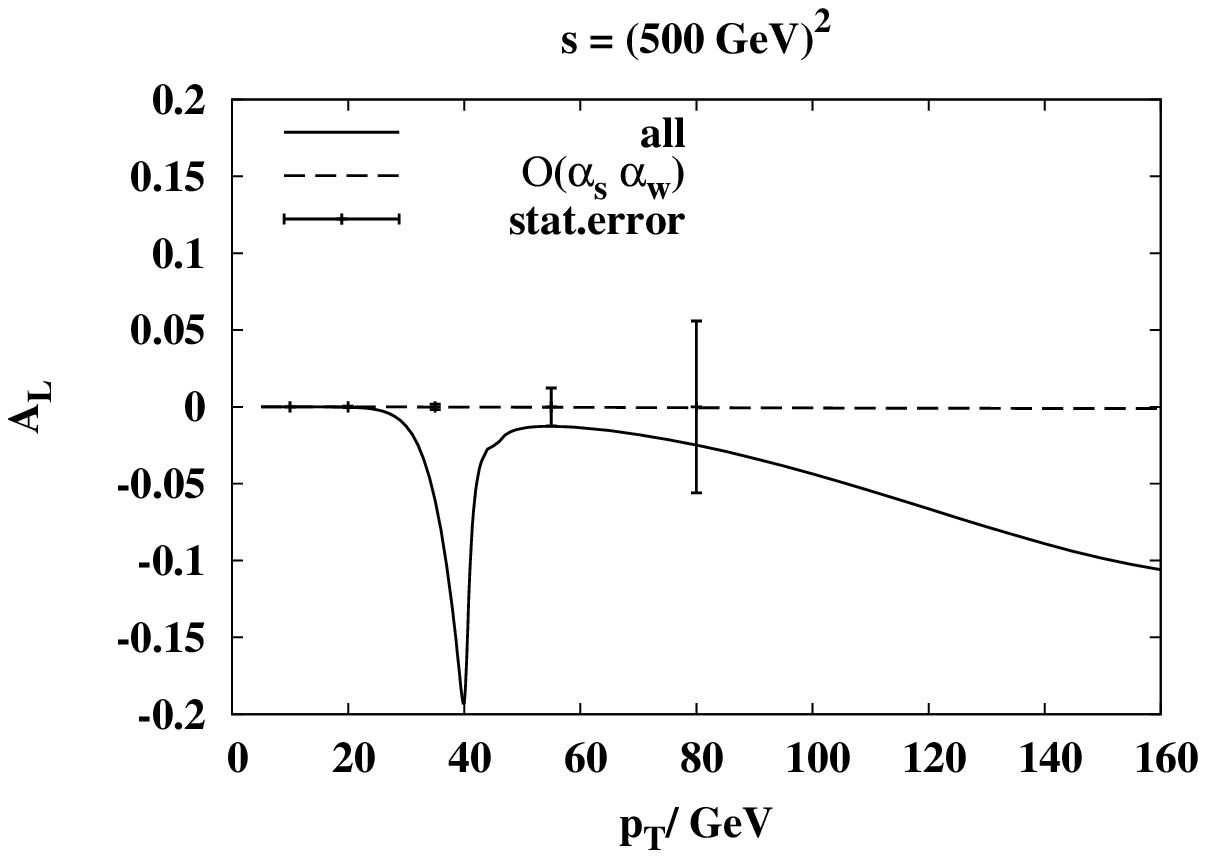}}
\centerline{(b)} 
\end{minipage} \\ 
\end{tabular} 
\caption{Single-spin asymmetry $A_L$ at LO for charm production. 
Dashed lines: contributions of 
${\cal O}(\alpha_S^2)+ {\cal O}(\alpha_S\alpha_W)$; solid lines: 
including also the contributions of ${\cal O}(\alpha_W^2)$. 
The indicated statistical error projections are estimated assuming an integrated 
luminosity of $100 \, \textrm{pb}^{-1}$ for $\sqrt{s} = 200 \, \textrm{GeV}$
and of $500 \, \textrm{pb}^{-1}$ for $\sqrt{s} = 500 \, \textrm{GeV}$, as well
as a beam polarization of $70\%$. They do not take  into account a charm
detection efficiency.} 
\label{f:charm} 
\end{center} 
\end{figure} 
Ignoring intrinsic charm in the nucleon, only two of the pure-QCD 
channels survive for charm production, namely $q \bar{q} 
\to c \bar{c}$ and $gg \to c\bar{c}$, leading to an enormously reduced 
unpolarized cross section in comparison to jet production.
Unfortunately, however, also the contribution of ${\cal O}(\alpha_S\alpha_W)$ 
becomes much smaller. The only partonic subprocesses for this
are $d \bar{d} ,\, s \bar{s} \to c \bar{c}$, which are either CKM-suppressed
or enter with the strange quark distribution.
Therefore, the main contributions to $\sigma_{{\mathrm{pol}}}$ are
of ${\cal O}(\alpha_W^2)$, as shown in Fig.~\ref{f:charm}, similar
to the lepton decay channel. Nonetheless, one can see from the figure 
that the asymmetry $A_L$ integrated over the charm's
pseudo-rapidity $|\eta|\leq 1$ reaches up to 
about $15 \, \%$ for $\sqrt{s} = 200 \, \textrm{GeV}$, and up to  
$10 \, \%$ for $\sqrt{s} = 500 \, \textrm{GeV}$. The pronounced peaks at 
both energies are generated by $s$-channel production of the weak gauge 
bosons, most notably the $W$. The statistical errors of the asymmetry, 
which have been estimated under our earlier assumptions without including
a charm detection efficiency, are reasonably small in the peak-region.
In Fig.~\ref{f:charm_eta}, $A_L$ is shown as a function of $\eta$, for 
different values of $p_T$. The effects can become of the order of $50 \%$, 
i.e., as large as in the case of the production of a lepton pair. 
Charm production probes in particular the large-$x$ $u$-quark 
distribution in the proton, and the $\bar{d}$ distribution at moderately
low $x$.  

\section{Summary and Conclusions} 
We have studied parity-violating single spin asymmetries in proton-proton 
scattering with hadronic final states for RHIC kinematics.  
In general, measuring such parity-violating observables can provide new,
complementary information on the quark helicity distributions of the nucleon.
Inclusive jets are produced copiously, but suffer from a very
small spin asymmetry $A_L$ (typically below $1 \,\%$). Also, because
of the many partonic channels that contribute to jets, the information
on the polarized parton distributions from $A_L$ for jets is not as
clear-cut as for the usually considered lepton final state. Still, 
the jet $A_L$ should be accessible at RHIC with the projected luminosity
and might be a tool for an early detection of a parity-violating signal.
We have estimated the role of QCD higher-order effects by implementing
leading-logarithm threshold resummation and found that the cross sections
are increased by resummation, but that the asymmetry is decreased further.
For a charm final state, rates are reduced, but the asymmetry can become 
much larger.

\section*{Acknowledgments}
We thank J.C. Bourrely, P. Nadolsky, E. Sichtermann, and J. Soffer for useful 
discussions. We are grateful to P. Schweitzer for directing our 
interest to charm production. 
W.V.\ is grateful to the U.S.\ Department of Energy (contract no.\
DE-AC02-98CH10886).


\end{document}